\documentclass[lettersize,journal]{IEEEtran}

\usepackage{amsmath,amsfonts}
\usepackage{algorithm}
\usepackage{array}
\usepackage[caption=false,font=normalsize,labelfont=sf,textfont=sf]{subfig}
\usepackage{textcomp}
\usepackage{stfloats}
\usepackage{url}
\usepackage{verbatim}
\usepackage{graphicx}
\usepackage{cite}
\usepackage[table]{xcolor}
\usepackage{soul}
\usepackage{float}
\usepackage{threeparttable}
\usepackage{booktabs}
\usepackage{multirow}
\newcommand{\tool}{\texttt{Trident}}

\begin{document}


\title{Seeing is Believing: Vision-driven Non-crash Functional Bug Detection for Mobile Apps}

\author{Zhe Liu, Cheng Li, Chunyang Chen, Junjie Wang~\IEEEmembership{Member,~IEEE}, Mengzhuo Chen, Boyu Wu, \\ Yawen Wang, Jun Hu, Qing Wang~\IEEEmembership{Member,~IEEE}
\thanks{This work was supported in part by the National Natural
Science Foundation of China under Grant 62232016, Grant 62072442, and
Grant 62272445; in part by Youth Innovation Promotion Association Chinese
Academy of Sciences.}
\thanks{Zhe Liu, Cheng Li, Junjie Wang, Mengzhuo Chen, Boyu Wu, Yawen Wang, Jun Hu, and Qing Wang are with State Key Laboratory of Intelligent Game, Institute of Software Chinese Academy of Sciences, University of Chinese Academy of Sciences,
Beijing 100190, China (e-mail: junjie@iscas.ac.cn;
liuzhe2020@iscas.ac.cn; wq@iscas.ac.cn).}
\thanks{Chunyang Chen is with the Technical University of Munich, D-80333 Munich,
Germany (e-mail: chunyang.chen@monash.edu).}}

\markboth{Journal of \LaTeX\ Class Files,~Vol.~14, No.~8, August~2021}%
{Shell \MakeLowercase{\textit{et al.}}: A Sample Article Using IEEEtran.cls for IEEE Journals}





\maketitle

\begin{abstract}
Mobile app GUI (Graphical User Interface) pages now contain rich visual information, with the visual semantics of each page helping users understand the application logic. However, these complex visual and functional logic present new challenges to software testing. Existing automated GUI testing methods, constrained by the lack of reliable testing oracles, are limited to detecting crash bugs with obvious abnormal signals. Consequently, many non-crash functional bugs, ranging from unexpected behaviors to logical errors, often evade detection by current techniques.
While these non-crash functional bugs can exhibit visual cues that serve as potential testing oracles, they often entail a sequence of screenshots, and detecting them necessitates an understanding of the operational logic among GUI page transitions, which is 
challenging traditional techniques.
Considering the remarkable performance of Multimodal Large Language Models (MLLM) in visual and language understanding, this paper proposes {\tool}, a novel vision-driven, multi-agent collaborative automated GUI testing approach for detecting non-crash functional bugs. 
It comprises three agents: Explorer, Monitor, and Detector, to guide the exploration, oversee the testing progress, and spot issues.
We also address several challenges, i.e., align visual and textual information for MLLM input, achieve functionality-oriented exploration, and infer test oracles for non-crash bugs, to enhance the performance of functionality bug detection. 
We evaluate {\tool} on 590 non-crash bugs and compare it with 12 baselines, it can achieve more than 14\%-112\% and 108\%-147\% boost in average recall and precision compared with the best baseline.
The ablation study further proves the contribution of each module.
Moreover, {\tool} identifies 43 new bugs on Google Play, of which 31 have been fixed. 
\end{abstract}


\section{Introduction}
\label{sec_introduction}

With the proliferation of mobile applications, their impact on everyday life and commercial operations has grown significantly. To meet the increasing demands for advanced functionalities, app features have become more complex, inevitably leading to the introduction of bugs. Existing research primarily focuses on crash bugs due to their clear test oracles ~\cite{mirzaei2016reducing,yang2018static,yang2013grey,zeng2016automated,mao2016sapienz,su2017guided,dong2020time,gu2019practical,wang2020combodroid}. However, there is a substantial number of non-crash functional bugs that remain underexplored.
According to our statistics on over 100,000 GitHub issue reports from 2021 to 2024, approximately 61.3\% describe non-crash functional bugs, with 47.7\% of them tagged as ``important''. These non-crash functional bugs can significantly impair functionality and even lead to disastrous consequences in various sectors and everyday routines.
For instance, on February 29, 2024, a software payment functionality bug caused widespread malfunctions in self-service payment systems at gas stations across New Zealand, resulting in severe repercussions~\cite{NewZealand}.

\IEEEpubidadjcol

\begin{figure*}
\centering
\vspace{0.4in}
\includegraphics[width=17.9cm]{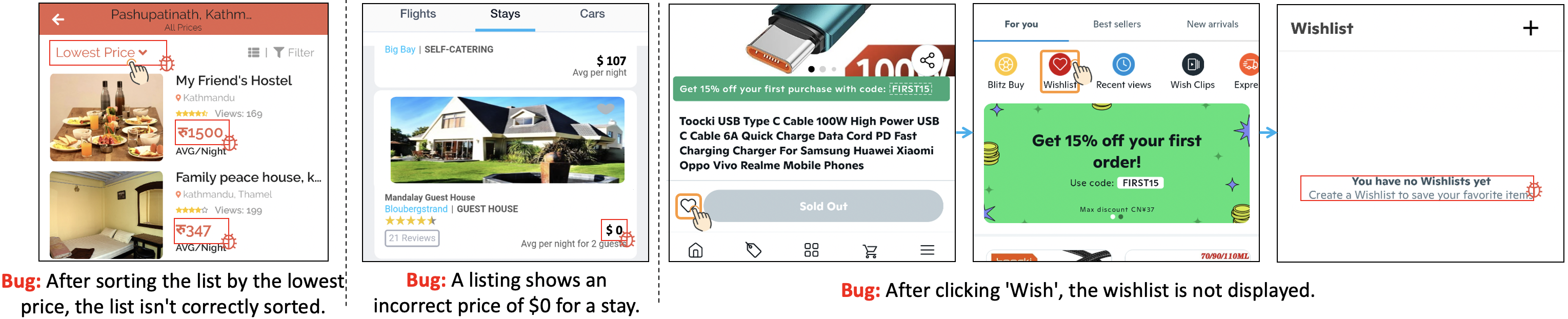}
\caption{Example of bugs with visual cues} 
\label{fig:bug-example}
\vspace{-0.1in}
\end{figure*}

On the other hand, we have observed that many non-crash functional bugs reveal clear visual cues. 
While some of these cues are obvious, such as component occlusions and text overlaps as investigated in previous studies~\cite{liu2020owl,liu2022Nighthawk}, more often they involve issues with the visual logic of the page. 
For instance, in the first bug shown in Fig. \ref{fig:bug-example}, the page is supposed to display items sorted by the lowest price, but categories with lower prices appear at the end instead. 
Similarly, in the third bug, clicking ``wish'' fails to display the corresponding item on the wishlist page. 
These bugs necessitate not only a semantic understanding of the GUI page content but also a comprehension of the overall operational logic underlying these GUI page transitions, making automated detection significantly challenging. 
Only human testers with both sharp eyes and knowledge from extensive training can identify those bugs.

However, human testing is time-consuming and suffers from inherent scalability and high-cost issues.
Although researchers begin to explore detecting non-crash or functional bugs with visual information, most of them are typically designed to identify specific types of bugs (e.g., display~\cite{liu2020owl}, game glitches~\cite{chen2021glib}).
What makes it worse is that all of them require either a set of rules that are easy to be out of date, or a large amount of training dataset which takes much human effort for annotation.
This specialization leaves a gap in the testing landscape for a more generalized, lightweight and intelligent approach that can adapt to the diverse challenges of mobile app testing especially for non-crash bugs.

Thanks to the emerging Multimodal Large Language Model (MLLM)~\cite{bubeck2023sparks}, it is trained on the ultra-large-scale corpus and potentially able to understand the GUI pages as well as the information changes along with the page transition. 
Inspired by this, we propose {\tool}, a novel vision-driven, multi-agent collaborative automated GUI testing approach for detecting non-crash functional bugs. 
It comprises three agents: Explorer, Monitor, and Detector. 
The Explorer Agent navigates through the app, captures view hierarchies and screenshots, and guides the exploration towards diverse GUI pages while focusing on the app’s functionalities. 
The Monitor Agent supervises the testing process, records the exploration history, and triggers the detector agent at the appropriate time. 
The Detector Agent identifies potential functional bugs by examining whether there are any issues in the logical transitions that occur during GUI page changes.


While the design of {\tool}'s three agents is intuitive, improving their effectiveness in revealing functional bugs presents several challenges. 
The first challenge is aligning visual and textual information for MLLM input. 
We develop a screenshot annotation method to guide the MLLM to better understand the GUI page, along with an alignment method that integrates text properties with visual context to enable the MLLM to capture comprehensive information. 
The second challenge involves achieving functionality-oriented exploration. 
We design a method to infer and abstract the tested functionality from detailed exploration sequences, and use the abstracted history when communicating with MLLM to avoid exceeding token limits and enhance the detection of functional bugs.
The third challenge is inferring test oracles for non-crash bugs. 
We design a mechanism to determine when to trigger the detector agent, and develop the functionality-aware chain-of-thought method to enable the MLLM to first explicitly infer oracles and then detect functional bugs based on these inferences, potentially reducing hallucinations in MLLM's decision-making process.

We evaluate the effectiveness of {\tool} on 590 non-crash bugs. Compared with 12 common-used and state-of-the-art baselines, {\tool} can achieve more than 14\%-112\% and 108\%-147\% boost in average recall and precision compared with the best baseline, resulting in 50\%-72\% precision and 42\%-52\% recall. To further understand the role of each sub-module of the approach, we conduct ablation experiments to further demonstrate its effectiveness.
We also evaluate the usefulness of {\tool} by detecting unseen non-crash bugs in real-world apps from Google Play. 
Among 187 apps, {\tool} detects 43 new crash bugs with 31 of them being fixed and 12 of them being confirmed by developers.

The contributions of this paper are as follows:
\begin{itemize}

\item The pioneering work to formulate the non-crash functional bugs detection problem to an interactive multimodal question \& answering task, enabling the MLLM to identify bugs by comprehending GUI visual information and functional logic.

\item A vision-driven automated GUI testing approach {\tool}\footnote{We release the source code, dataset, and experimental results on our website \url{https://github.com/testtest2024-art/Trident}. \label{github}} which includes a vision-driven explorer, process-aware monitor, and function-aware detector to identify non-crash functional bugs through MLLM.

\item Effectiveness and usefulness evaluation of the {\tool} in real-world apps with practical bugs detected and confirmed (Section \ref{sec_results_RQ1} and \ref{sec_results_RQ4}).
\end{itemize}

\begin{figure}[htb]
\centering
\includegraphics[width=8.7cm]{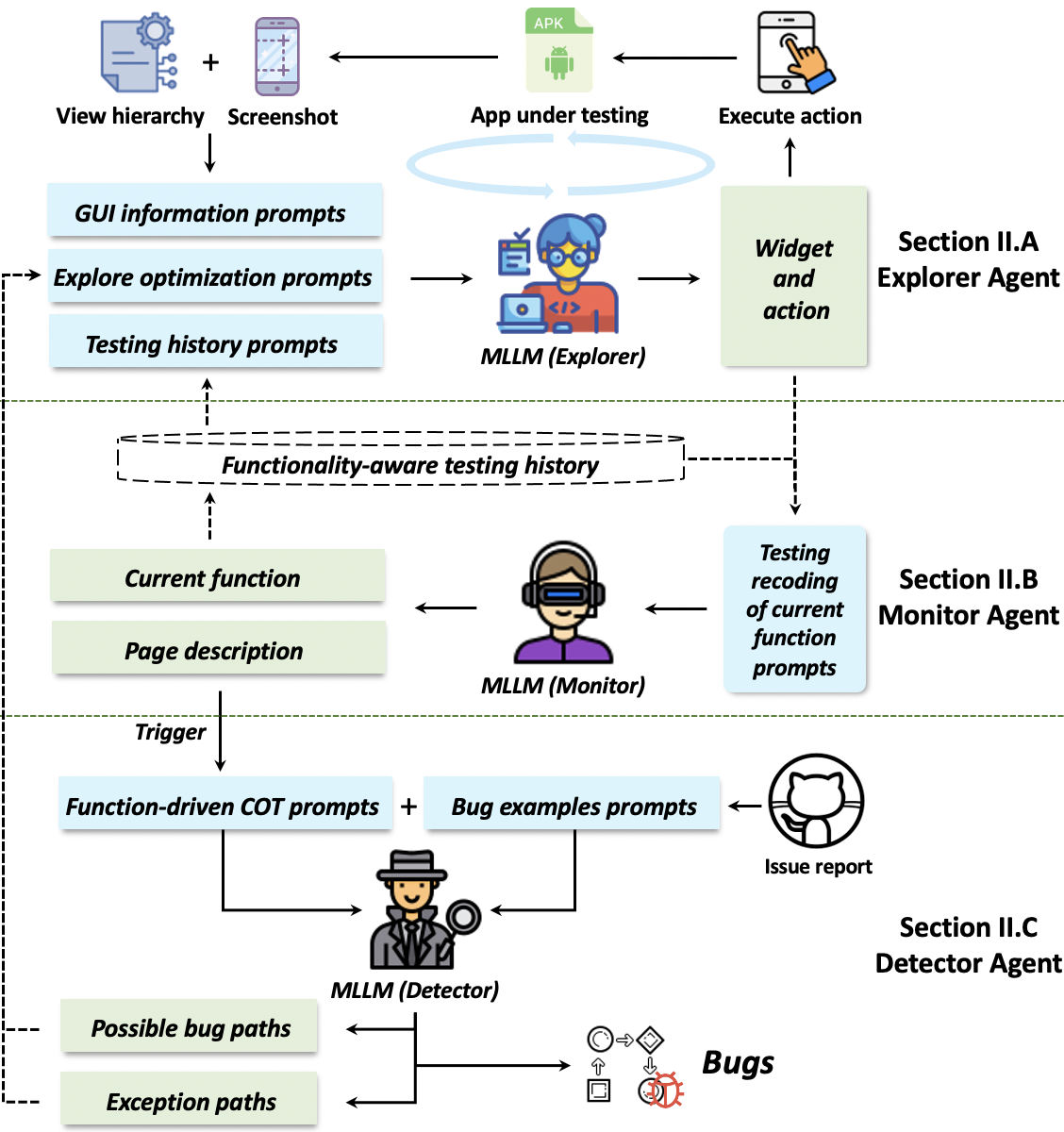}
\caption{Overview of {\tool}}
\label{fig:overview}
\vspace{-0.1in}
\end{figure}

\section{Approach}
\label{sec_approach}
This paper proposes {\tool}, a novel vision-driven, multi-agent collaborative automated GUI testing approach for detecting non-crash functional bugs. 
Inspired by manual GUI testing, where human testers navigate through the app, observe the testing process, and identify bugs through visual and logical analysis, {\tool} comprises three agents: Explorer, Monitor, and Detector, as demonstrated in Fig. \ref{fig:overview}.

The Explorer Agent navigates through the app, captures view hierarchies and screenshots, and guides the exploration towards diverse GUI pages while focusing on the app's functionalities.
The Monitor Agent supervises the testing process, records the exploration history, and triggers the detector agent at the appropriate time.
The Detector Agent identifies potential functional bugs by examining whether there are any issues in the logical transitions that occur during GUI page changes.

While the design of {\tool}'s three agents is intuitive, how to make them more effective in revealing functional bugs is difficult. 
Below, we outline three specific challenges.

\textbf{Challenge 1: Aligning visual and text for MLLM input.}
Although MLLMs can directly process images and plan explorations based on screenshots, their outputs are limited to text. 
This means that MLLMs cannot directly mark the next actionable widgets on a screenshot; they must rely on textual descriptions. 
To cope with this, we propose an alignment method that integrates text properties with visual context (see Section \ref{subsec_approach_Explore}). This allows the MLLM to have a more comprehensive understanding of the GUI, thereby better guiding subsequent exploration.
Furthermore, screenshots often contain numerous elements. 
If fed directly to the MLLM, it may overlook critical details. 
To mitigate this, we design a screenshot annotation method, guiding the MLLM to pay attention to different types of actionable widgets, and also resolve the issue of overlapping annotations caused by the density of widgets on the GUI page.


\textbf{Challenge 2: Functionality-oriented exploration. }
During app testing, a sequence of actions and accompanying page transitions is generated. These actions focus on the operational level of the app, making it difficult for LLM to plan exploration paths from a functional perspective, thus impacting the detection of functional bugs.
To address this, we develop a method to infer and abstract the current functionality from detailed exploration sequences (in Section \ref{subsec_approach_Monitor}). 
It helps in two ways: 1) avoids exceeding token limits when interacting with the LLM; 2) enables the exploration more focusing on the functionality aspect for identifying functional bugs.

\textbf{Challenge 3: Inferring test oracle. }
Accurately obtaining test oracles is a time-consuming and labor-intensive task, which has long been a limiting factor in the development of automated testing~\cite{li2019humanoid,li2017droidbot,su2017guided,pan2020reinforcement,dong2020time}. In GUI testing, it requires determining what the current GUI page should be like after each performed action. This not only involves information from a single page (e.g., missing text) but also considers a series of previous pages and actions (e.g., an item added not appearing on the list page), which is quite challenging. 
To address this, we let the Monitor Agent trigger the Detector Agent at the end of each functionality exploration
(see Section \ref{subsec_approach_Monitor}). This allows the Detector Agent to assess functionality by analyzing a manageable sequence of actions and states, thus focusing on the functionality and providing accurate inferences. 
Additionally, we develop the functionality-aware Chain-of-Thought (COT) to enable the MLLM to first explicitly infer oracles and then detect functional bugs based on these inferences, potentially reducing the hallucinations in MLLM's decision-making process (in Section \ref{subsec_approach_Detector_COT}).


\subsection{\textbf{Vision-driven Explorer Agent}}
\label{subsec_approach_Explore}
The \textbf{Explorer Agent} serves as the foundation of {\tool}, which visually inspects a GUI page, understands its semantics, and conducts the exploration. 
Before the MLLM, existing automated GUI testing approaches~\cite{li2017droidbot,li2019humanoid,moran2018mdroid+,su2017guided,pan2020reinforcement} typically relied on view hierarchy files to extract attribute information (text, coordinates, actions, etc.) of app widgets to capture the current context and determine which widgets should be acted upon. 
This can be inefficient as it may lose the structural semantics of the GUI when converting screenshots to text. 
Thanks to the MLLM, it allows us to use screenshot images as input directly. However, relying solely on images can also be limited due to the concise nature of GUI text and icons. 
Therefore, we design a method to align visual and textual information to enhance the MLLM's understanding of the GUI page semantics.

\subsubsection{\textbf{Alignment between Text and Image}}
\label{subsec_approach_Alignment}
For the \textbf{text}, we extract three types of information: app details, current GUI page context, and the contained widgets. The app information is obtained from the \textit{AndroidManifest.xml} file, while page and widget information are extracted from the view hierarchy file of the current GUI page using UIAutomator~\cite{uiautomator}. Initially, we gather information such as the \{App name\}, \{Activities\} and \{Activity name\} to provide the MLLM with a general understanding of the app's functionalities. Subsequently, we extract page information and widget information including fields such as ``text'', ``hint-text'', ``resource-id'', ``class'', ``clickable'', ``long-clickable'', ``checkable'', ``scroll'', and ``bounds'', which facilitate the MLLM in suggesting appropriate actions related to these widgets.


\begin{figure}[htb]
\centering
\vspace{-0.1in}
\includegraphics[width=8.7cm]{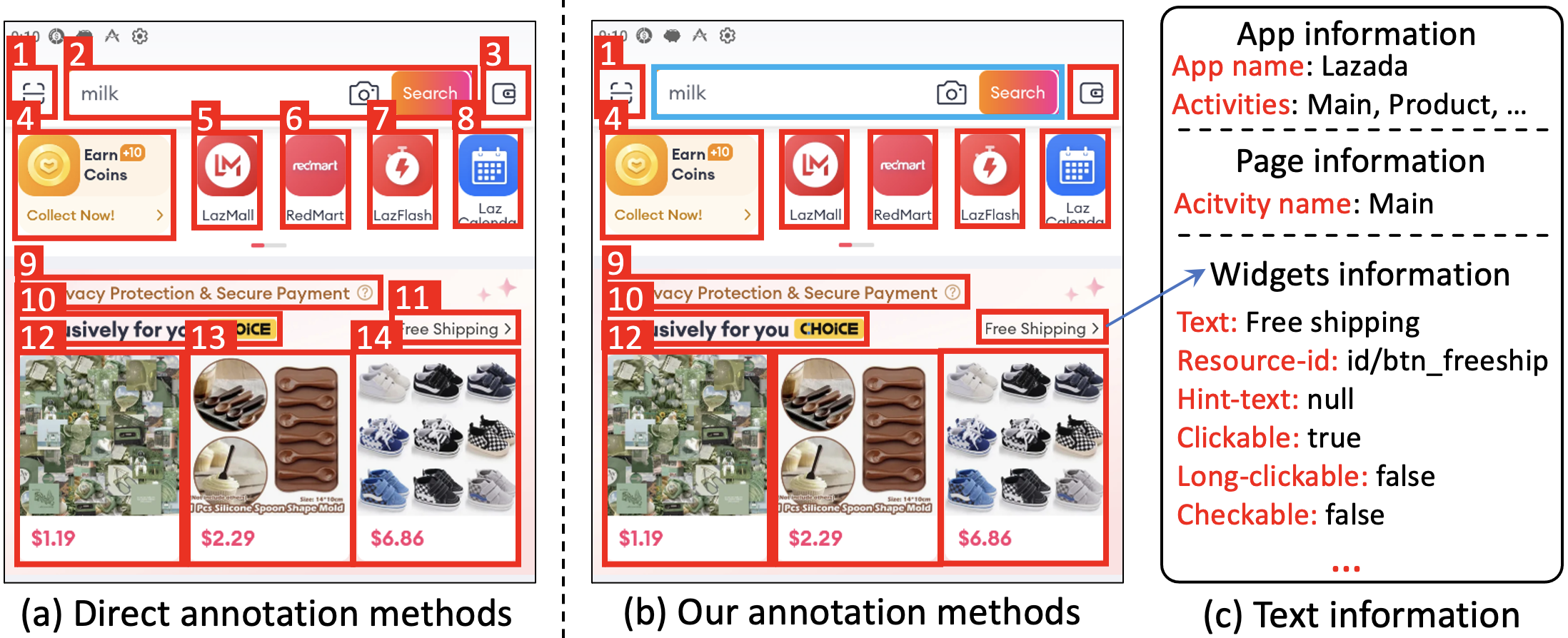}
\caption{Example of alignment text with image}
\label{fig:example-screenshot}
\vspace{-0.1in}
\end{figure}

\begin{figure}[htb]
\centering
\includegraphics[width=8.7cm]{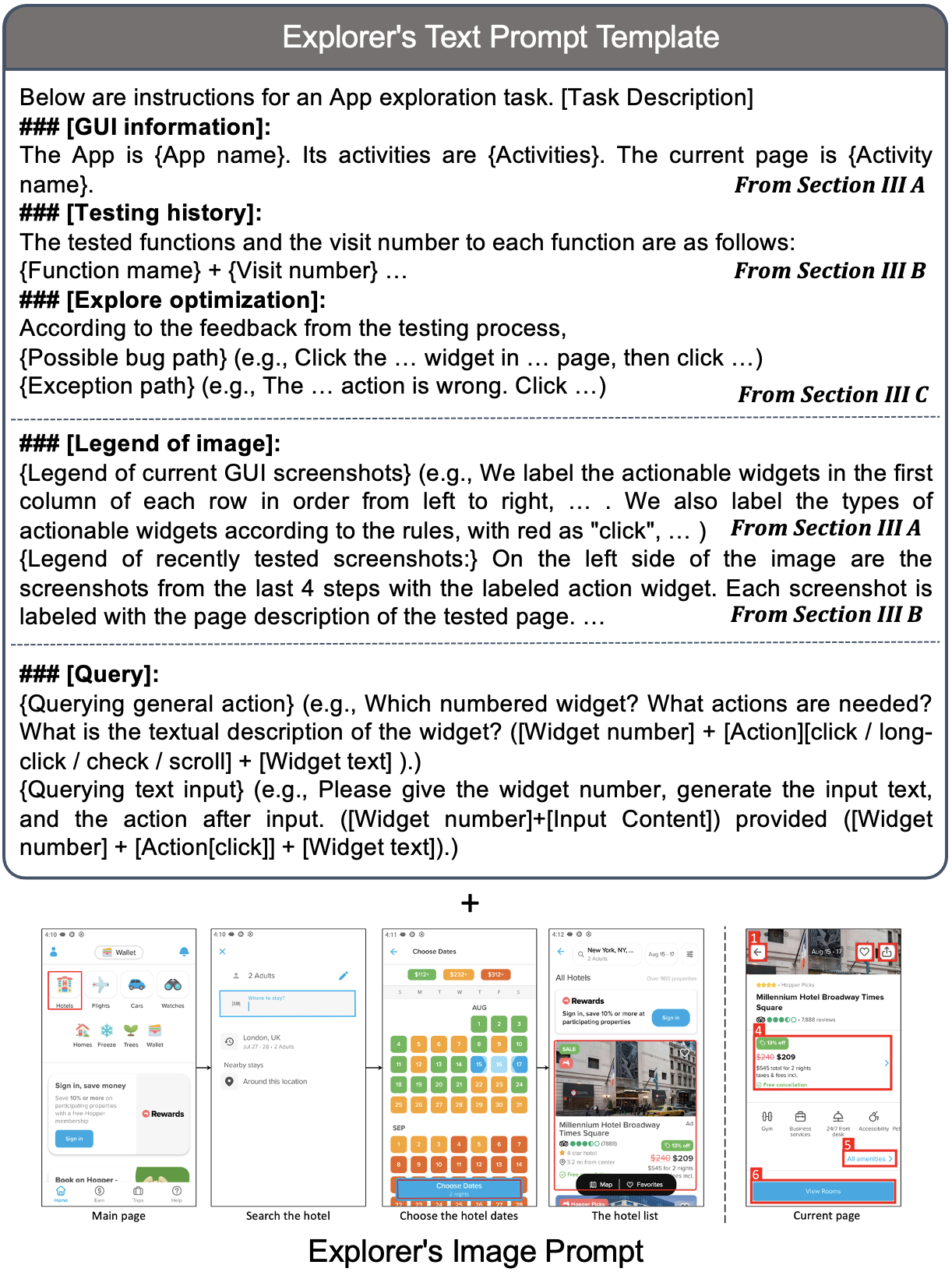}
\caption{Example of explorer prompt}
\label{fig:explore-prompt}
\vspace{-0.1in}
\end{figure}

For the \textbf{image}, we design a screenshot annotation method to guide the MLLM's attention towards actionable widgets, as illustrated in Fig. \ref{fig:example-screenshot}. 
{\tool} supports five basic operational actions: click, input, long-click, check, and scroll, while the less standardized actions such as drag-and-drop or multi-finger gestures are excluded, following existing studies~\cite{mao2016sapienz,su2017guided,dong2020time,gu2019practical,wang2020combodroid}. 
Bounding boxes are drawn around each actionable widget based on their coordinates from the ``bounds'' field, with different colors indicating different action types. For instance, a red bounding box indicates a ``clickable'' widget, a blue box indicates an ``EditText'' widget requiring text input, and other colors are used for long-clickable, checkable, and scrollable widgets. If a widget supports multiple actions, a green bounding box is used, with specific actions described in the accompanying text. Inspired by the screen reader~\cite{zhang2021screen}, we label each widget with numbers from top to bottom and left to right, marking only the first widget of each row to reduce the overlapping annotations and letting the MLLM infer others. 
To help the MLLM better understand the annotation information on the image, we also provide a \{legend of image\} in the prompt as shown in Fig. \ref{fig:explore-prompt}.

We associate each annotated widget in the screenshot with the extracted GUI text information. In prompts related to text and image information (details in Fig. \ref{fig:explore-prompt}), detailed alignment information is omitted due to input length constraints of the MLLM. Instead, this alignment is used to verify whether the MLLM suggests interactive widgets according to the correct logical process, akin to an explicit chain-of-thought mechanism. When querying general actions or text input, we request the widget ID from the screenshot and the corresponding widget text. If these sources do not align, we infer potential errors in the MLLM's decision-making and prompt it to retry. To prevent continuous errors and infinite loops, we limit retries to three attempts before triggering a fallback strategy.

\subsubsection{\textbf{Explorer Prompt}}

Fig. \ref{fig:explore-prompt} demonstrates the explorer prompt. 
It is composed of the GUI information and screenshots as described in this section, and the testing history outlining the dynamic testing process provided by the Monitor agent (in Section \ref{subsec_approach_Monitor}), as well as the explore optimization feedback from the Detector agent (in Section \ref{subsec_approach_Detector}). 
It queries the MLLM about the interactive widget and the action that needs to be taken.

\subsection{\textbf{Process-aware Monitor Agent}}
\label{subsec_approach_Monitor}

The \textbf{Monitor Agent} acts as the central control unit of {\tool}, overseeing the test progress, abstracting the tested functionalities history for the Explorer Agent, and triggering the Detector Agent at appropriate times. 
It continuously captures the fine-grained exploration sequence of the app, including view hierarchies and screenshots of each explored GUI page, and the related actions on those pages.
Based on this information, it generates two views of testing histories: a higher-level text view in the form of explored functionalities and a detailed snapshot of testing records in an image view.
During the testing process, these views, which represent transformed testing histories, are provided to the Explorer Agent, which addresses the token limit for LLM's inputs while also enhancing the understanding of the testing process.
During the monitoring process, if the Monitor Agent determines that the current functionality is still under exploration, it instructs the Explorer Agent to continue. Once it identifies that a new functionality starts to test, the Monitor Agent pauses the Explorer Agent and triggers the Detector Agent to analyze potential bugs in the previously tested functionality.

\begin{figure}[htb]
\centering
\vspace{-0.1in}
\includegraphics[width=8.6cm]{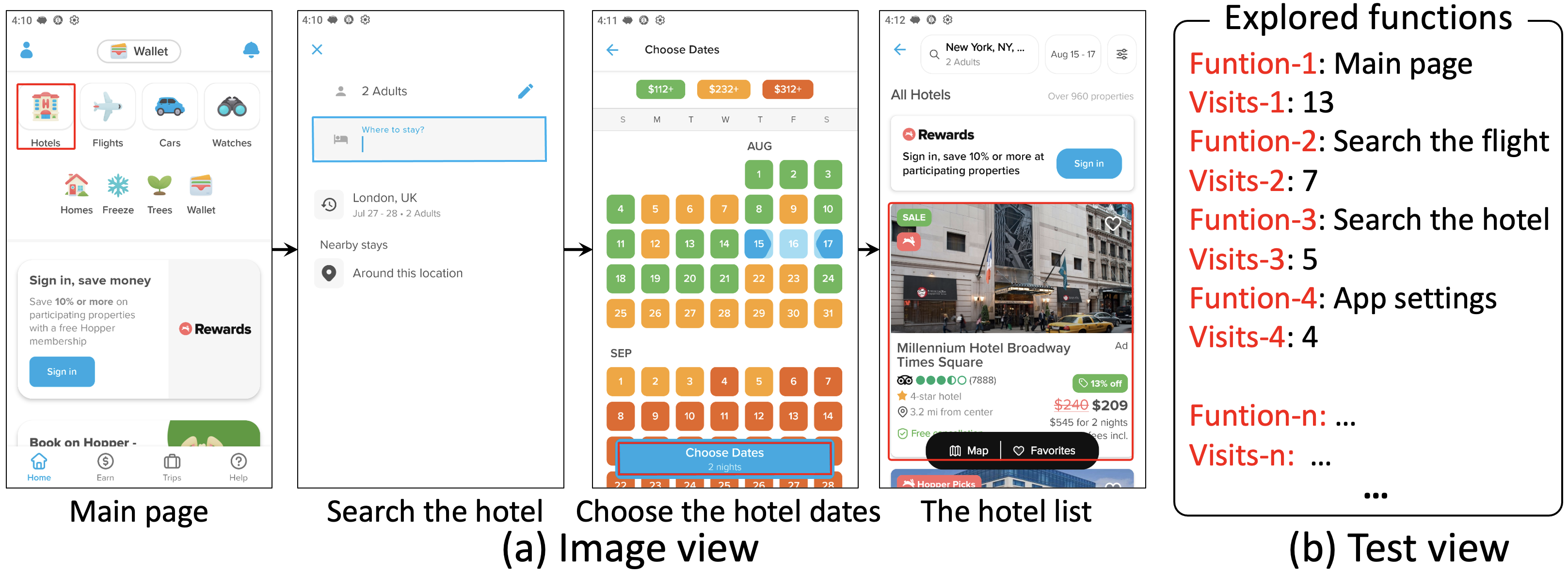}
\caption{Example of function-based history}
\label{fig:example-history}
\vspace{-0.1in}
\end{figure}

\subsubsection{\textbf{Function-based Testing History Construction}}
\label{subsec_approach_Monitor_construction}

To facilitate exploration of the app’s functionality, it is important to know the testing process, as the functions are usually accomplished by following a sequence of GUI pages. 
We maintain a complete testing history database, including textual descriptions and screenshots, and offer two views supporting the follow-up exploration and bug detection, as demonstrated in Fig. \ref{fig:example-history}.

The text view depicts the higher-level catalog of testing progress, organized according to the functionalities.
A catalog of tested functionalities \{Functionality list\} with their \{Visit number\} (visit counts) and \{Functionality status\} (e.g., fully tested, partially tested). This assists the MLLM in avoiding redundant exploration and identifying untouched functionalities. During testing, the MLLM infers the current functionality being tested through a functionality inquiry prompt, enabling the collection of functionality-specific statistics.

The image view offers a detailed snapshot of the testing records through screenshots, providing a granular view of the testing process. Each GUI page is marked with widgets or actions performed, and screenshots are associated with page descriptions. 
This ensures alignment between visual and textual information and supports thorough exploration and bug detection.

\begin{figure}[htb]
\centering
\includegraphics[width=8.7cm]{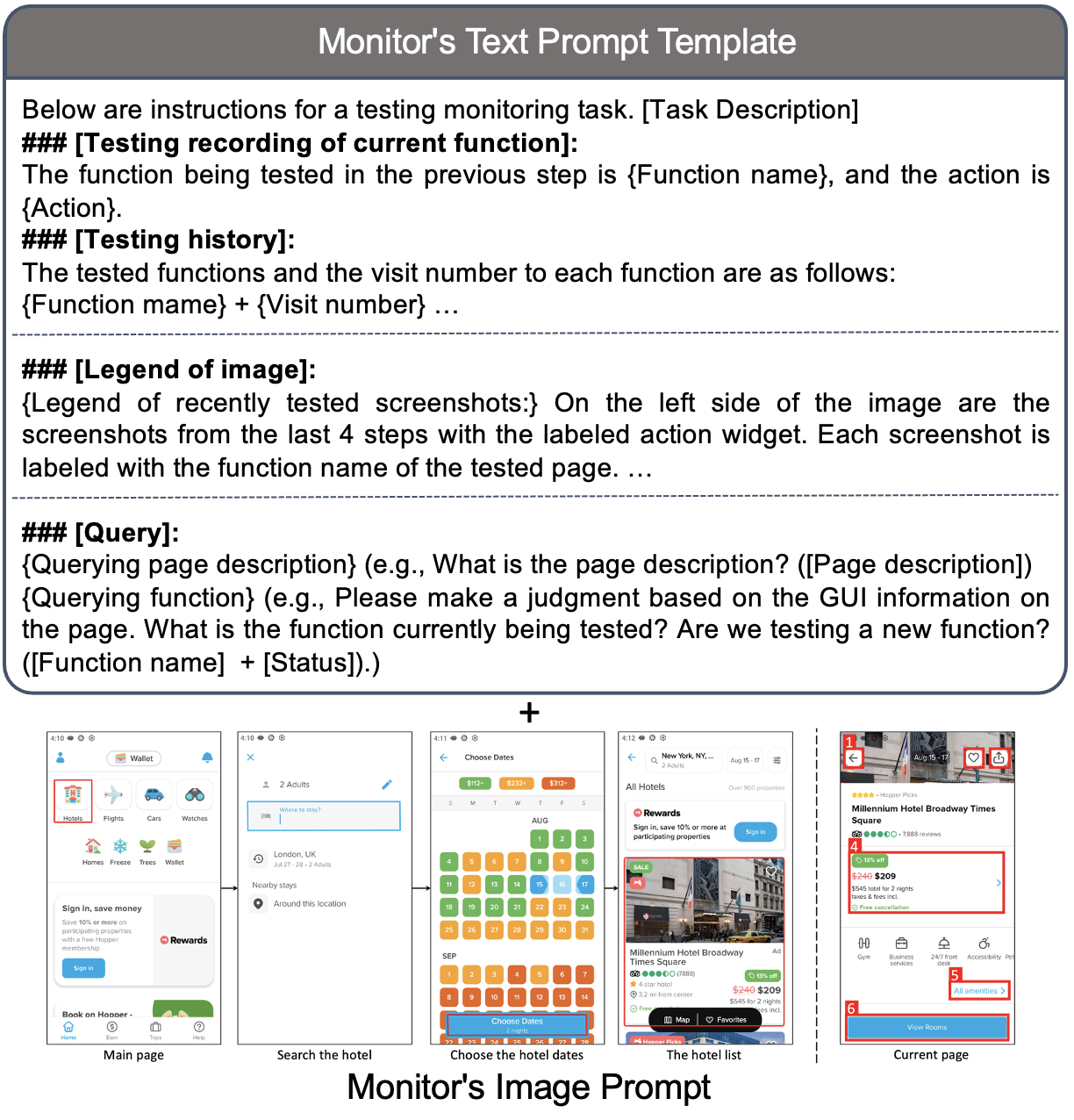}
\caption{Example of monitor prompt}
\label{fig:Monitor-prompt}
\vspace{-0.1in}
\end{figure}

\subsubsection{\textbf{Monitor Prompt}}

Fig. \ref{fig:Monitor-prompt} demonstrates the monitor prompt.
It inputs the testing recording of the current function and the whole testing history, in the form of both text and image, and queries the MLLM about the GUI page description, currently tested function and its status. 



\subsection{\textbf{Function-aware Detector Agent}}
\label{subsec_approach_Detector}

The \textbf{Detector Agent} is the vigilant observer of {\tool} that identifies potential functional bugs by analyzing the logical transitions and interactions within the GUI pages.
Upon activation by the Monitor Agent, the Detector Agent 
first examines the execution sequence of functionality to identify logical inconsistencies for bug detection, in which we design the functionality-driven Chain-of-Thought strategy and enrich the prompt with examples to enhance the performance.
The Detector then analyzes the sequence to uncover any potential bug-triggering operations that have not yet been explored. 
It also identifies any unusual or abnormal operations within the execution sequence that might influence the exploration performance.
These results are fed back to the Explorer Agent,  which will integrate these insights into the follow-up exploration.
This iterative cycle ensures that the GUI testing is thorough and efficient, enhancing the likelihood of uncovering non-crash functional bugs while systematically addressing any anomalies detected during the process.

\begin{figure}[htb]
\centering
\includegraphics[width=8.7cm]{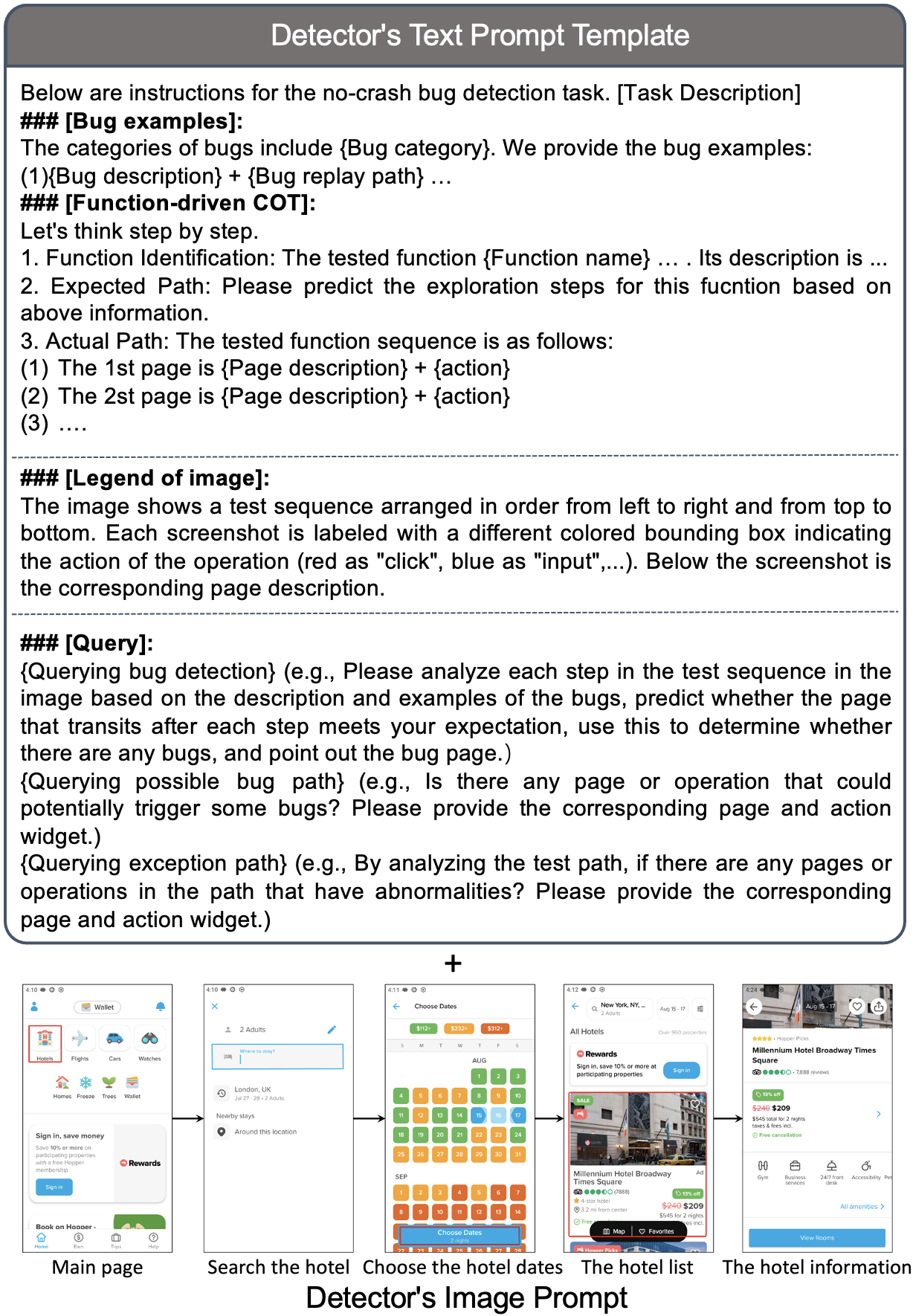}
\caption{Example of detector prompt}
\label{fig:Detector-prompt}
\vspace{-0.1in}
\end{figure}

\subsubsection{\textbf{Functionality-driven Chain of Thought (COT)}}
\label{subsec_approach_Detector_COT}
We design a functionality-driven Chain of Thought (COT) method, which guides the LLM through a structured, sequential analysis of the testing functionality. 
This ensures that each step is carefully evaluated, reducing the likelihood of hallucination and enhancing the overall effectiveness of the bug detection process.

The MLLM first analyzes the functionality being tested, the expected path, and the actual path taken during testing. This involves the following steps:
Functionality Identification: The model identifies the functionality being tested, including its name and description. This helps the model understand the specific functionality it needs to analyze.
Expected Path: The model predicts the exploration steps for the functionality based on its description. This step sets a benchmark for the expected behavior.
Actual Path: The model reviews the actual sequence of actions and states encountered during testing. Each page and corresponding action are listed sequentially, allowing the model to compare this path against the expected behavior.
The prompt also includes an annotated image of the test sequence, with each screenshot labeled according to the type of action (e.g., click, input). This visual representation helps the model correlate the text information with the GUI elements.

\subsubsection{\textbf{Enriching Detector Prompt with Examples}}
\label{subsec_approach_Prompt-tuning}
It is usually difficult for MLLM to perform as well on domain-specific tasks as ours, and common practice would be employing the in-context learning schema to boost the performance. 
It provides the MLLM with examples to demonstrate what the instruction is, which enables the MLLM to better understand the task. 
Following the schema, along with the prompt for the detector, as described in Fig. \ref{fig:Detector-prompt}, we additionally provide the MLLM with examples of the non-crash logical bugs. 

To achieve this, we first build a basic example dataset of non-crash functional bugs from the issue reports of open-source mobile apps and continuously enlarge it. 
For each data instance, it records the bug description, bug screenshot and natural language described bug reproduction path which facilitates the MLLM understanding of what the non-crash functional bugs are. 
We select 200 bugs with the highest quality and complete information as the example data.
We also enrich the example dataset with the newly emerged non-crash functional bugs that run on various apps.  

Although examples can provide intuitive guidance to the LLM in accomplishing a task, excessive examples might mislead the LLM and cause the performance to decline. 
Therefore, we design a retrieval-based example selection method to choose the most suitable examples (i.e., the bug page's description is most similar to the currently tested GUI page's description) for MLLM.
The similarity comparison is based on the activity name of the tested page. 
Similar to previous works~\cite{liu2023chatting}, We use Word2Vec~\cite{Word2Vec} to encode the context information of each activity into a sentence embedding, and calculate the cosine similarity between the activity and each data instance in the example dataset. 
We choose the top-K data instance with the highest similarity score and set K as 5.

\subsubsection{\textbf{Detector Prompt}}

Fig. \ref{fig:Detector-prompt} demonstrates the detector prompt. 
It consists of the bug examples (as shown in Section \ref{subsec_approach_Prompt-tuning}), functionality-driven COT (as shown in Section \ref{subsec_approach_Detector_COT}) analyzing the functionality being tested, and the real exploration sequence in the form of screenshots. It queries the MLLM about the bug detection, possible bug path, and exception path.

\subsection{\textbf{Implementation}}
\label{subsec_approach_Implement}
In our implementation, we utilize the GPT-4 Vision model\footnote{\url{https://platform.openai.com/docs/models/gpt-4-turbo-and-gpt-4}} (gpt-4-vision-preview) to handle both textual and visual information during the testing process. It obtains the view hierarchy file of the current GUI page through UIAutomator~\cite{uiautomator} to extract text information of the input widgets.
{\tool} is implemented as a fully automated GUI testing tool, which uses or extends the following tools: VirtualBox and the Python pyvbox~\cite{pyvbox} for running and Android Debug Bridge (ADB)~\cite{Adb} for interacting with the app under test.

To manage the token limit constraints of the GPT-4V API, we combine multiple screenshots from the testing sequences into a single large image. This approach allows us to input comprehensive visual information cohesively, ensuring the model processes all necessary contexts together. On average, each API call consumes approximately 2,000 to 3,000 tokens. We maintain concise representations of the testing history by using functionality names and summaries, reducing the token count. 
Additionally, when the limit is exceeded, we delete the earlier test records until it satisfies the constraints.

To enhance efficiency and reduce costs, we merge the prompts for Explorer and Monitor into a single request. 
This integrated prompt simultaneously queries the MLLM about the current exploration state (for explorer) and the tested functionalities (for monitor), which makes the implementation more efficient and cost-effective.  

\section{Experiment Design}
\label{sec_Experiment_Design}

\begin{itemize}
\item \textbf{RQ1: (Coverage and bugs detection performance)} How effective of {\tool} in testing coverage and bug detection? 
\end{itemize}

For RQ1, we first present some general views of {\tool} for coverage of the explorer and bug detection of the detector. Then we compare {\tool} with commonly-used and state-of-the-art baseline approaches. 

\begin{itemize}
\item \textbf{RQ2: (Ablation Study)} 
What is the contribution of the (sub-) modules of {\tool} for coverage and bug detection performance?
\end{itemize}

For RQ2, We conduct ablation experiments to evaluate the impact of each (sub-) module on the performance.

\begin{itemize}
\item \textbf{RQ3: (Usefulness Evaluation)} How does our proposed {\tool} work in real-world situations?
\end{itemize}

For RQ3, we use {\tool} to automatically explore the app, and detect unseen non-crash functional bugs, and issue the detected bugs to the development team.

\subsection{\textbf{Datasets}}
\label{subsec_experiment_dataset}
\textbf{For RQ1-RQ2}, the experimental dataset comes from three sources. 
The first is from the Odin and RegDroid datasets \cite{xiong2023empirical,wang2022detecting} (We name them \textit{baseline dataset}), which contains 131 non-crash functional bugs that can be reproduced in our experimental environment. 

Meanwhile, we notice that these apps are updated between 2018 and 2021, and some of them are now incompatible with the latest systems. 
To further investigate the effectiveness of {\tool} on more recently updated apps and new types of bugs (the \textit{baseline dataset} only with 5 categories such as ``Content related issues'', ``Structure related issues'', ``Incorrect interaction logic'', etc.), we collect a second dataset (\textit{Github dataset}) following similar procedures as the baseline dataset, which contain a total of 9 categories of bugs\textsuperscript{\ref{github}}, such as ``Page loading failure'', ``Feedback missing'', ``Numerical calculation errors'', ``Data operation failures'', ``Navigation and link errors'', .etc.
In detail, we crawl the 50 most popular apps of each category from Google Play~\cite{Googleplay}, and we keep the ones with at least one update after June. 2024, resulting in 437 apps in 13 Google Play categories~\cite{Googleplay}. 
Then we manually review each issue report and app, and filter them according to the following criteria: (1) the app wouldn't constantly crash on the emulator; (2) it can run all baselines; (3) UIAutomator~\cite{uiautomator} can obtain the view hierarchy file and its GUI screenshot; (4) the issue report involve non-crash bug and it can be manually reproduced for validation; (5) the app is not used in the bug example dataset. 
Please note that we follow the name of the app to ensure that there is no overlap between the datasets.
Finally, 83 apps with 387 non-crash bugs remain for further experiments.


To avoid potential data leakage issues caused by the use of MLLM and further analyze the bug detection capability of {\tool}, we construct a third dataset (\textit{Injection dataset}) through bug injection, which can ensure that MLLM has not learned about these bugs in the open-source repositories. 
Specifically, we obtain the source code of the second dataset apps from GitHub and manually verify whether the app could conduce the bug injection by modifying the code through repackaging. 
Among them, 33 apps can run normally after bug injection. 
We conduct the injection based on the bug description, bug repair code, and bug root cause from GitHub, and implement it following the existing bug injection approach \cite{ghaleb2020effective,bhattacharya2013empirical,hu2014efficiently}. Finally, 33 apps with 72 non-crash bugs remain for further experiments.

The three datasets totally have 590 non-crash bugs for comprehensive evaluation of the effectiveness of {\tool}.

\begin{table*}[htb]
\vspace{0.1in}
\caption{Result of coverage and bug detection (RQ1)}
\vspace{-0.05in}
\label{tab:RQ1-result}
\centering
\scriptsize
\begin{tabular}{p{1.8cm}<{\centering} | p{0.45cm}<{\centering} | p{0.45cm}<{\centering} | p{0.45cm}<{\centering} | p{0.45cm}<{\centering} | p{0.45cm}<{\centering} | p{0.45cm}<{\centering} | p{0.45cm}<{\centering} | p{0.45cm}<{\centering} | p{0.45cm}<{\centering} | p{0.45cm}<{\centering} | p{0.45cm}<{\centering} | p{0.45cm}<{\centering} | p{0.45cm}<{\centering} | p{0.45cm}<{\centering} | p{0.45cm}<{\centering} | p{0.45cm}<{\centering} | p{0.45cm}<{\centering} | p{0.45cm}<{\centering}}
\toprule
\multirow{2}*{\textbf{Method}} & \multicolumn{6}{c|}{\textbf{Baseline data}} & \multicolumn{6}{c|}{\textbf{GitHub data}} & \multicolumn{6}{c}{\textbf{Bug injection data}} \cr 
& \textbf{Act} & \textbf{Cod} & \textbf{Rep} & \textbf{Bug} & \textbf{P} & \textbf{R} & \textbf{Act} & \textbf{Cod} & \textbf{Rep} & \textbf{Bug} & \textbf{P} & \textbf{R} & \textbf{Act} & \textbf{Cod} & \textbf{Rep} & \textbf{Bug} & \textbf{P} & \textbf{R}\\ 
\midrule
OwlEyes+M & 0.21 & 0.18 & 77 & 10  & 0.13  & 0.08  & 0.33 & 0.28 & 140 & 62  & 0.44  & 0.16  & 0.35 & 0.31 & 96 & 13 & 0.14  & 0.18 \\ 
\rowcolor{gray!15}
NightHawk+M & 0.21 & 0.18 & 63 & 13  & 0.21  & 0.10  & 0.33 & 0.28 & 136 & 62  & 0.46  & 0.16  & 0.35 & 0.31 & 60 & 15 & 0.25  & 0.21 \\ 
DiffDroid+M & 0.21 & 0.18 & 61 & 15  & 0.25  & 0.11  & 0.33 & 0.28 & 133 & 55  & 0.41  & 0.14  & 0.35 & 0.31 & 46 & 12 & 0.26  & 0.17 \\ 
\rowcolor{gray!15}
dVermin+M & 0.21 & 0.18 & 53 & 4  & 0.08  & 0.03  & 0.33 & 0.28 & 98 & 28  & 0.29  & 0.07  & 0.35 & 0.31 & 33 & 11 & 0.33  & 0.15 \\ 
\midrule
OwlEyes+A  & 0.29 & 0.27 & 92 & 18  & 0.20  & 0.14  & 0.43 & 0.37 & 161 & 85  & 0.53  & 0.22  & 0.42 & 0.4 & 93 & 10 & 0.11  & 0.14 \\ 
\rowcolor{gray!15}
NightHawk+A  & 0.29 & 0.27 & 86 & 21  & 0.24  & 0.16  & 0.43 & 0.37 & 157 & 91  & 0.58  & 0.24  & 0.42 & 0.4 & 67 & 16 & 0.24  & 0.22 \\ 
DiffDroid+A  & 0.29 & 0.27 & 87 & 20  & 0.23  & 0.15  & 0.43 & 0.37 & 147 & 68  & 0.46  & 0.18  & 0.42 & 0.4 & 44 & 7 & 0.16  & 0.10 \\ 
\rowcolor{gray!15}
dVermin+A  & 0.29 & 0.27 & 67 & 6  & 0.09  & 0.05  & 0.43 & 0.37 & 101 & 31  & 0.31  & 0.08  & 0.42 & 0.4 & 35 & 9 & 0.26  & 0.13 \\ 
\midrule
OwlEyes+F  & 0.39 & 0.37 & 107 & 17  & 0.16  & 0.13  & 0.47 & 0.42 & 180 & 79  & 0.44  & 0.20  & 0.43 & 0.39 & 94 & 12 & 0.13  & 0.17 \\ 
\rowcolor{gray!15}
NightHawk+F  & 0.39 & 0.37 & 100 & 20  & 0.20  & 0.15  & 0.47 & 0.42 & 161 & 92  & 0.57  & 0.24  & 0.43 & 0.39 & 59 & 15 & 0.25  & 0.21 \\ 
DiffDroid+F  & 0.39 & 0.37 & 94 & 20 & 0.21  & 0.15  & 0.47 & 0.42 & 131 & 72 & 0.55  & 0.19  & 0.43 & 0.39 & 53 & 12 & 0.23  & 0.17 \\
\rowcolor{gray!15}
dVermin+F  & 0.39 & 0.37 & 41 & 8  & 0.20  & 0.06  & 0.47 & 0.42 & 90 & 35  & 0.39  & 0.09  & 0.43 & 0.39 & 67 & 8 & 0.12  & 0.11 \\ 
\midrule
OwlEyes+H  & 0.38 & 0.35 & 97 & 12  & 0.12  & 0.09  & 0.48 & 0.44 & 157 & 79  & 0.50  & 0.20  & 0.43 & 0.39 & 93 & 14 & 0.15  & 0.19 \\ 
\rowcolor{gray!15}
NightHawk+H  & 0.38 & 0.35 & 88 & 16  & 0.18  & 0.12  & 0.48 & 0.44 & 146 & 79  & 0.54  & 0.20  & 0.43 & 0.39 & 88 & 13 & 0.15  & 0.18 \\ 
DiffDroid+H  & 0.38 & 0.35 & 75 & 17  & 0.23  & 0.13  & 0.48 & 0.44 & 143 & 57  & 0.40  & 0.15  & 0.43 & 0.39 & 57 & 15 & 0.26  & 0.21 \\ 
\rowcolor{gray!15}
dVermin+H  & 0.38 & 0.35 & 33 & 5  & 0.15  & 0.04  & 0.48 & 0.44 & 82 & 40  & 0.49  & 0.10  & 0.43 & 0.39 & 41 & 8 & 0.20  & 0.11 \\ 
\midrule
OwlEyes+G  & 0.47 & 0.43 & 99 & 17  & 0.17  & 0.13  & 0.51 & 0.47 & 162 & 79  & 0.49  & 0.20  & 0.47 & 0.41 & 95 & 14 & 0.15  & 0.19 \\ 
\rowcolor{gray!15}
NightHawk+G  & 0.47 & 0.43 & 95 & 18  & 0.19  & 0.14  & 0.51 & 0.47 & 157 & 96  & 0.61  & 0.25  & 0.47 & 0.41 & 68 & 12 & 0.18  & 0.17 \\ 
DiffDroid+G  & 0.47 & 0.43 & 85 & 16  & 0.19  & 0.12  & 0.51 & 0.47 & 142 & 76  & 0.54  & 0.20  & 0.47 & 0.41 & 47 & 9 & 0.19  & 0.13 \\ 
\rowcolor{gray!15}
dVermin+G  & 0.47 & 0.43 & 37 & 6  & 0.16  & 0.05  & 0.51 & 0.47 & 75 & 47  & 0.63  & 0.12  & 0.47 & 0.41 & 42 & 9 & 0.21  & 0.13 \\ 
\midrule
iFixdataloss  & 0.33 & 0.3 & 21 & 8  & 0.38  & 0.06  & 0.48 & 0.43 & 101 & 45  & 0.45  & 0.12  & 0.43 & 0.39 & 41 & 10 & 0.24  & 0.14 \\ 
\rowcolor{gray!15}
SetDroid  & 0.31 & 0.27 & 22 & 8  & 0.36  & 0.06  & 0.48 & 0.43 & 164 & 64  & 0.39  & 0.17  & 0.43 & 0.39 & 47 & 11 & 0.23  & 0.15 \\ 
\midrule
Genie  & 0.32 & 0.31 & 55 & 12  & 0.22  & 0.09  & 0.48 & 0.43 & 123 & 46  & 0.37  & 0.12  & 0.43 & 0.39 & 43 & 8 & 0.19  & 0.11 \\ 
\rowcolor{gray!15}
Odin  & 0.33 & 0.3 & 57 & 22  & 0.39  & 0.17  & 0.48 & 0.43 & 107 & 59  & 0.55  & 0.15  & 0.43 & 0.39 & 57 & 11 & 0.19  & 0.15 \\ 
\midrule
Lint  &  -   &  -   & 1 & 1  & 1.00  & 0.01  &  -   &  -  & 54 & 30  & 0.56  & 0.08  &  -  &  -  & 2 & 2 & 1.00  & 0.03 \\ 
\rowcolor{gray!15}
LiveDroid  &  -   &  -   & 1 & 1  & 1.00  & 0.01  &  -   &  -  & 58 & 24  & 0.41  & 0.06  &  -  &  -  & 1 & 1 & 1.00  & 0.01 \\ 
\midrule
GPT-4  & 0.39 & 0.31 & 55 & 12  & 0.22  & 0.09  & 0.47 & 0.45 & 123 & 48  & 0.39  & 0.12  & 0.43 & 0.39 & 35 & 8 & 0.23  & 0.11 \\ 
\rowcolor{gray!15}
AppAgent & 0.41 & 0.38 & 47 & 6  & 0.13  & 0.05  & 0.44 & 0.43 & 113 & 12  & 0.11  & 0.03  & 0.39 & 0.37 & 31 & 4 & 0.13  & 0.06 \\ 
\textbf{{\tool}}  & \textbf{0.57} & \textbf{0.55} & \textbf{109} & \textbf{55} & \textbf{0.50}  & \textbf{0.42}  & \textbf{0.65} & \textbf{0.62} & \textbf{278} & \textbf{201} & \textbf{0.72}  & \textbf{0.52}  & \textbf{0.61} & \textbf{0.59} & \textbf{67} & \textbf{47} & \textbf{0.70}  & \textbf{0.65} \\ 

\bottomrule
\end{tabular}
\vspace{0.05in}
\begin{tablenotes}
\scriptsize
\item \textbf{\textit{Notes:}} ``Act'' is activity coverage, ``Cod'' is Code coverage. ``Rep'' is number of bugs reported by tools, ``Bug'' is true bug in the reported bug, ``P'' is precision and ``R'' is the recall. ``+M'' means it with Monkey, ``+A'' means it with APE, ``+F'' means it with Fastbot, ``+H'' means it with Hunmanoid, and ``+G'' means it with GPTDroid.
\end{tablenotes}
\end{table*}

\textbf{For RQ3,} we further evaluate the usefulness of {\tool} in detecting unseen bugs. 
We randomly selected 187 popular apps based on app type and download number.
We use the same configurations as the previous experiments. 
Once a bug is spotted, we create an issue report by describing the bug, and report them to the app development team through the issue reporting system or email.

\subsection{\textbf{Baselines}}
\label{subsec_experiment_baseline}
To demonstrate the advantage of {\tool}, we compare it with 12 common-used and state-of-the-art GUI testing techniques.

There are 4 mobile app display issue detection techniques. 
OwlEyes~\cite{liu2020owl} and NightHawk~\cite{liu2022Nighthawk} use deep learning to detect UI display issues (e.g., overlapping texts, component occlusion). 
DVermin~\cite{su2022metamorphosis} detects scaling issues by comparing GUI pages under different page settings. 
DiffDroid~\cite{fazzini2017automated} compares the GUI pages of an app on two different devices to find compatibility issues. 
The input of these techniques are screenshots/visual hierarchy files and need to be integrated with automated testing approaches to fetch the required information during the exploration. 
We respectively pair them with five popular automated GUI testing tools, i.e., random-based tool Monkey~\cite{Monkey}, model-based tools APE~\cite{gu2019practical} and Fastbot~\cite{cai2020fastbot}, learning-based tool Humanoid~\cite{li2019humanoid} and LLM-based GPTDroid~\cite{liu2023chatting}. 

We utilize 6 baselines of logical bug detection for mobile apps. 
iFixdataloss~\cite{guo2022detecting} is the most recent work of finding data loss issues. 
SetDroid~\cite{sun2023characterizing} uses metamorphic testing to find system setting-related functional bugs. 
Genie~\cite{su2021fully} uses the metamorphic relation of independent view property to find non-crash functional bugs, while Odin~\cite{wang2022detecting} uses the implicit knowledge of ``bugs as deviant behavior'' to find non-crash functional bugs. 
Android Lint~\cite{lint} is a popular static analysis tool for finding different issues in Android apps based on predefined rules. 
LiveDroid~\cite{farooq2020livedroid} is also a static analysis tool for detecting data loss issues.

We also employ 2 MLLM-based methods as our baselines. AppAgent~\cite{yang2023appagent} is a novel MLLM-based framework that enables agents to interact with apps through a simplified action space, mimicking human-like interactions such as tapping and swiping. 
We modify AppAgent by adding a query in its prompt after each test interaction to check for any detected bugs on the page. Additionally, we include GPT-4 as another baseline. We provided GPT-4 with annotated screenshots of the app. GPT-4 was then tasked with generating actions based on these screenshots and detecting any potential bugs.

\subsection{\textbf{Experimental Setup}}
We deploy the baselines and our approach on a 64-bit machine and evaluate them on Google Android emulators. 
Following common practice~\cite{su2017guided,dong2020time,wang2020combodroid}, we registered separate accounts for each bug that requires login and wrote the login scripts, and during testing reset the account data before each run to avoid possible interference.
To ensure fair and reasonable use of resources, we set up the running time of each tool in one app to 50 minutes, which is widely used in other GUI testing studies~\cite{dong2020time,wang2020combodroid,li2017droidbot,cai2020fastbot}.
We run each tool 3 times and take the average value as the result to mitigate potential bias~\cite{dong2020time,wang2020combodroid,li2017droidbot,cai2020fastbot}.

\begin{table*}[htb]
\caption{Contribution of different sub-modules (RQ2)}
\vspace{-0.05in}
\label{tab:RQ2-1}
\centering
\scriptsize
\begin{tabular}{p{2.7cm}<{\centering} | p{0.4cm}<{\centering} | p{0.4cm}<{\centering} | p{0.4cm}<{\centering} | p{0.4cm}<{\centering} | p{0.4cm}<{\centering} | p{0.4cm}<{\centering} | p{0.4cm}<{\centering} | p{0.4cm}<{\centering} | p{0.4cm}<{\centering} | p{0.4cm}<{\centering} | p{0.4cm}<{\centering} | p{0.4cm}<{\centering} | p{0.4cm}<{\centering} | p{0.4cm}<{\centering} | p{0.4cm}<{\centering} | p{0.4cm}<{\centering} | p{0.4cm}<{\centering} | p{0.4cm}<{\centering}}
\toprule
\multirow{2}*{\textbf{Sub-modules}} & \multicolumn{6}{c|}{\textbf{Baseline data}} & \multicolumn{6}{c|}{\textbf{GitHub data}} & \multicolumn{6}{c}{\textbf{Bug injection data}} \cr 
& \textbf{Act} & \textbf{Cod} & \textbf{Rep} & \textbf{Bug} & \textbf{P} & \textbf{R} & \textbf{Act} & \textbf{Cod} & \textbf{Rep} & \textbf{Bug} & \textbf{P} & \textbf{R} & \textbf{Act} & \textbf{Cod} & \textbf{Rep} & \textbf{Bug} & \textbf{P} & \textbf{R}\\ 
\midrule
\textbf{{\tool}}  & \textbf{0.57} & \textbf{0.55} & \textbf{109} & \textbf{55} & \textbf{0.50}  & \textbf{0.42}  & \textbf{0.65} & \textbf{0.62} & \textbf{278} & \textbf{201} & \textbf{0.72}  & \textbf{0.52}  & \textbf{0.61} & \textbf{0.59} & \textbf{67} & \textbf{47} & \textbf{0.70}  & \textbf{0.65} \\ 
\midrule
w/o Screenshot annotation & 0.41 & 0.48 & 55 & 27 & 0.49  & 0.21  & 0.41 & 0.39 & 101 & 49 & 0.49  & 0.13  & 0.38 & 0.36 & 39 & 15 & 0.38  & 0.21 \\ 
\rowcolor{gray!15}
w/o GUI information & 0.49 & 0.47 & 69 & 33 & 0.48  & 0.25  & 0.52 & 0.50 & 222 & 129 & 0.58  & 0.33  & 0.49 & 0.45 & 52 & 31 & 0.60  & 0.43 \\ 
w/o Testing history & 0.39 & 0.37 & 43 & 21 & 0.49  & 0.16  & 0.42 & 0.39 & 131 & 53 & 0.40  & 0.14  & 0.44 & 0.42 & 43 & 19 & 0.44  & 0.26 \\ 
\rowcolor{gray!15}

w/o Optimization & 0.47 & 0.45 & 39 & 16 & 0.41  & 0.12  & 0.44 & 0.42 & 87 & 38 & 0.44  & 0.10  & 0.42 & 0.41 & 42 & 17 & 0.40  & 0.24 \\ 
w/o Image Legend & 0.38 & 0.36 & 27 & 12 & 0.44  & 0.09  & 0.39 & 0.38 & 78 & 31 & 0.40  & 0.08  & 0.42 & 0.41 & 39 & 15 & 0.38  & 0.21 \\ 
\rowcolor{gray!15}
\midrule
w/o Test Recoding & 0.42 & 0.39 & 38 & 18 & 0.47  & 0.14  & 0.48 & 0.46 & 98 & 51 & 0.52  & 0.13  & 0.43 & 0.42 & 41 & 19 & 0.46  & 0.26 \\ 
\midrule
w/o Bug example & 0.53 & 0.52 & 47 & 9 & 0.19  & 0.07  & 0.59 & 0.58 & 51 & 17 & 0.33  & 0.04  & 0.54 & 0.52 & 38 & 5 & 0.13  & 0.07 \\ 
\rowcolor{gray!15}
w/o Function COT & 0.51 & 0.49 & 54 & 11 & 0.20  & 0.08  & 0.58 & 0.55 & 63 & 23 & 0.37  & 0.06  & 0.55 & 0.53 & 39 & 7 & 0.18  & 0.10 \\
\bottomrule
\end{tabular}
\end{table*}

\subsection{\textbf{Evaluation Metrics}}

We measure the performance from coverage and bug detection, respectively corresponding to the explorer and detector in {\tool}.
For coverage, we obtain the activity coverage and code coverage~\cite{he2020textexerciser,liu2017automatic,arnatovich2018mobolic,wang2020combodroid,wang2021vet,li2019humanoid}, in which we treat the activities defined in the \textit{AndroidManifest.xml} file of an Android app as the whole set of  activities~\cite{pan2020reinforcement,Android,su2017guided}.

For bug detection, following existing studies~\cite{su2021fully,wang2022detecting}, we report the detected bugs by {\tool}, true bugs among the detected bugs.
Since we know the entire set of bugs, we also calculate the precision and recall of the bug detection. 

\section{Results and Analysis}
\subsection{\textbf{Coverage and Bug Detection Performance (RQ1)}}
\label{sec_results_RQ1}
We present the coverage and bug detection performance of {\tool} and baselines, in terms of the baseline dataset, GitHub dataset and the bug injection dataset in Table \ref{tab:RQ1-result}.

We first present the coverage of {\tool}, which can cover far more activities and code than the baselines. In three datasets, the average activity coverage is between 57\% and 65\%, and the average code coverage is between 55\% and 62\%. 
It achieves between 21\% (0.57 vs. 0.47) to 27\% (0.65 vs. 0.51) activity coverage, and between 28\% (0.55 vs. 0.43) and 32\% (0.62 vs. 0.47) code coverage higher even compared with the best baseline (GPTDroid) across three datasets. 
This indicates the effectiveness of {\tool} in covering more activities and codes, thus bringing higher confidence to app quality, potentially uncovering more bugs.

The average precision and recall of {\tool} in three datasets achieve 50\%-72\%, and 42\%-65\%, respectively.
It achieves between 14\% (0.72 vs. 0.63) and 112\% (0.70 vs. 0.33) precision and between 108\% (0.52 vs. 0.25) and 147\% (0.42 vs. 0.17) recall higher even compared with the best baseline across three datasets. 
Please note that although some baseline methods can achieve higher precision, they detect fewer bugs. In terms of the number of detected bugs, {\tool} can still achieve a higher detection rate.
We also compare the similarities and differences of the bugs between the best baselines (Odin and NightHawk) and our approach, and the results show that all bugs detected by them are also detected by {\tool}. 
This indicates the effectiveness of {\tool} in detecting bugs and helps to ensure app quality. 

The advantages of {\tool} are due to three aspects. 
First, {\tool} designs the vision prompt which can better capture the GUI context, while the baselines mainly rely on the view hierarchy file in which the widgets might have meaningless ``text'' or ``resource-id'', hindering them from effectively understanding the GUI pages. 
Second, {\tool} plans the exploration path both considering the semantics of a single GUI page and the exploration history which can enable the function-aware exploration, while the baselines can hardly understand the inherent business logic of the app, resulting in aimless exploration. 
Third, {\tool} detects the bugs by understanding the semantics of a sequence of GUI pages, while the baselines can only detect specific types of bugs or have low detection accuracy.  

\begin{figure}
\centering
\includegraphics[width=8.6cm]{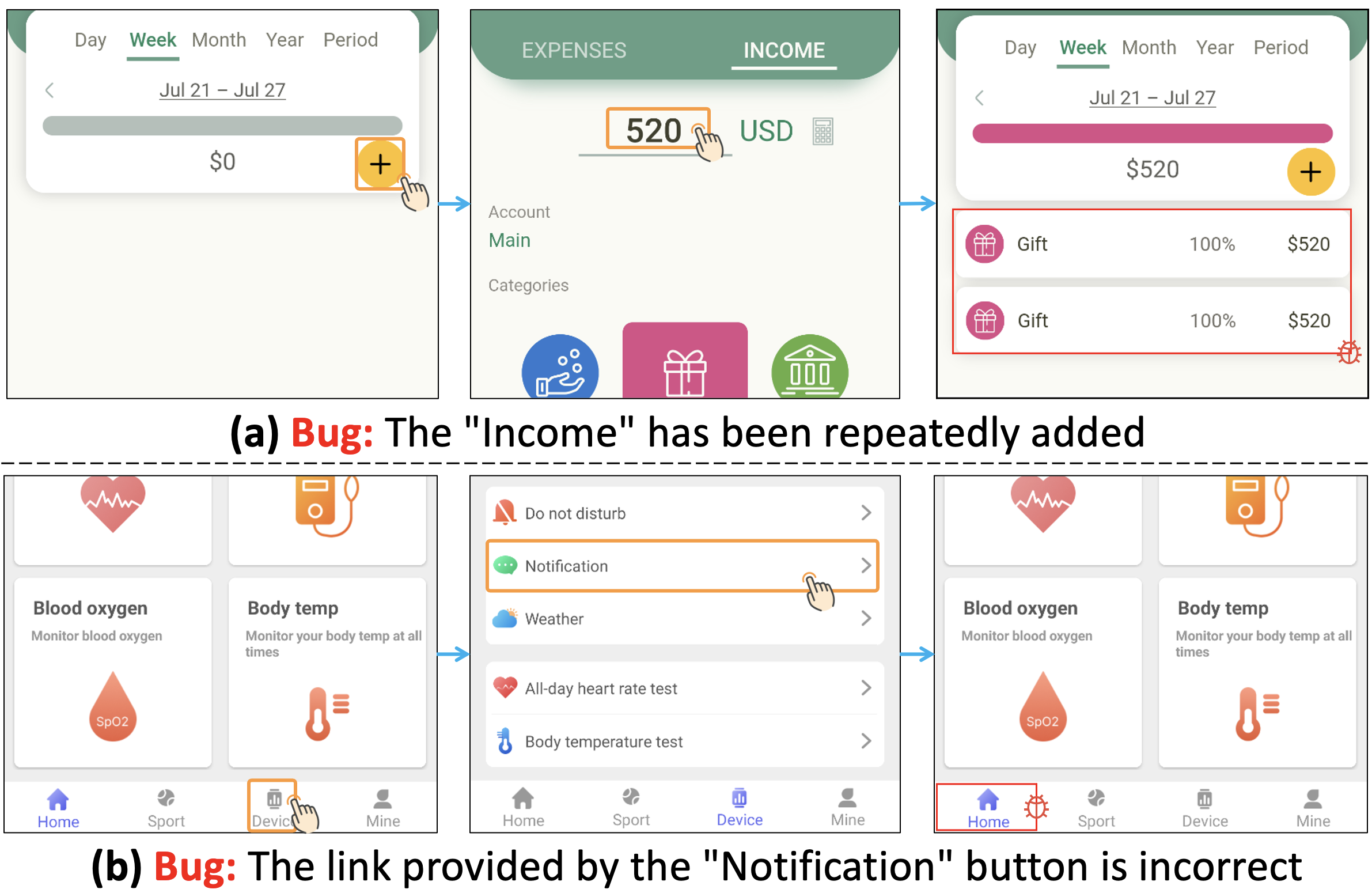}
\caption{Example of bugs detected by {\tool}} 
\label{fig:goodcase}
\vspace{-0.15in}
\end{figure}

Fig. \ref{fig:bug-example} and Fig. \ref{fig:goodcase} show some examples of bugs detected by {\tool}. These bugs belong to different types and are not detected by the baseline method. In Fig. \ref{fig:goodcase} (a), when the user adds income, enter \$520. After returning to the main page, the income was repeatedly displayed in the list, resulting in the same income being incorrectly recorded twice.
In Fig. \ref{fig:goodcase} (b), the user clicks on the ``Devices'' tab and then clicks on the ``Notifications''. As a result, the app returns to the main page instead of the correct notification page.
We can see that {\tool} is adept at determining whether the functional logic of the app is abnormal based on the semantics of the test path and GUI information.


\subsection{\textbf{Ablation Study (RQ2)}}
\label{sec_results_RQ2}

\subsubsection{\textbf{Contribution of Sub-modules}}
\label{subsub_results_RQ3_sub}
Table \ref{tab:RQ2-1} further demonstrates the performance of {\tool} and its 8 variants. We remove each sub-module of the {\tool} in Fig. \ref{fig:overview} separately, i.e., Screenshot annotation, GUI information, Testing history, Optimization, Image Legend, Test Recoding, Bug example and Function COT.
For removing the Screenshot annotation, We adopt the direct annotation method shown in Fig. \ref{fig:example-screenshot} (a). For removing GUI information, Testing history, and Optimization, we separately delete three corresponding parts from the prompt in Fig. \ref{fig:explore-prompt}. For removing the Image Legend, we delete the image legend in three prompts. For removing the Test Recoding, we delete it from the prompt in Fig. \ref{fig:Monitor-prompt}. For removing the Bug example and Function COT, we delete them from the prompt in Fig. \ref{fig:Detector-prompt}.

The experiment results (Table \ref{tab:RQ2-1}) demonstrate that removing any of the sub-modules would result in a noticeable performance decline, indicating the necessity and effectiveness of the designed sub-modules. 
Removing the bugs example (\textit{w/o-Bug example}) has the greatest impact on the performance, reducing the bug detection performance by 54\% (0.33 vs. 0.72) precision and 92\% (0.04 vs. 0.52) recall.
This indicates that by providing similar examples, the MLLM can quickly think out what should the non-crash bugs look like, which further indicates the demonstration can facilitate the MLLM in producing the required output.

We also notice that, the \textit{w/o-Screenshot annotation}, \textit{w/o-Testing history} and \textit{w/o-Image Legend}, the activity coverage has an impact, reducing the activity coverage by 37\% (0.41 vs. 0.65), 35\% (0.42 vs. 0.65) and 40\% (0.39 vs. 0.65). This further indicates that the alignment of text with images can help MLLM understand the structure and semantic information of GUI pages and make reasonable judgments. The history can help retain knowledge during testing and gain global viewpoints to reach uncovered areas. 

Besides, (\textit{w/o-Function COT}) would also largely influence the performance, and decrease the bug detection performance by 56\% (0.37 vs. 0.72) precision and 88\% (0.06 vs. 0.52) recall.
This indicates that through COT, MLLM can infer the possible path of the functionality based on its description, and then judge the testing path, improving MLLM's understanding of the functionality and bug detection performance.

\begin{figure}[htb]
\centering
\vspace{-0.1in}
\includegraphics[width=7.7cm]{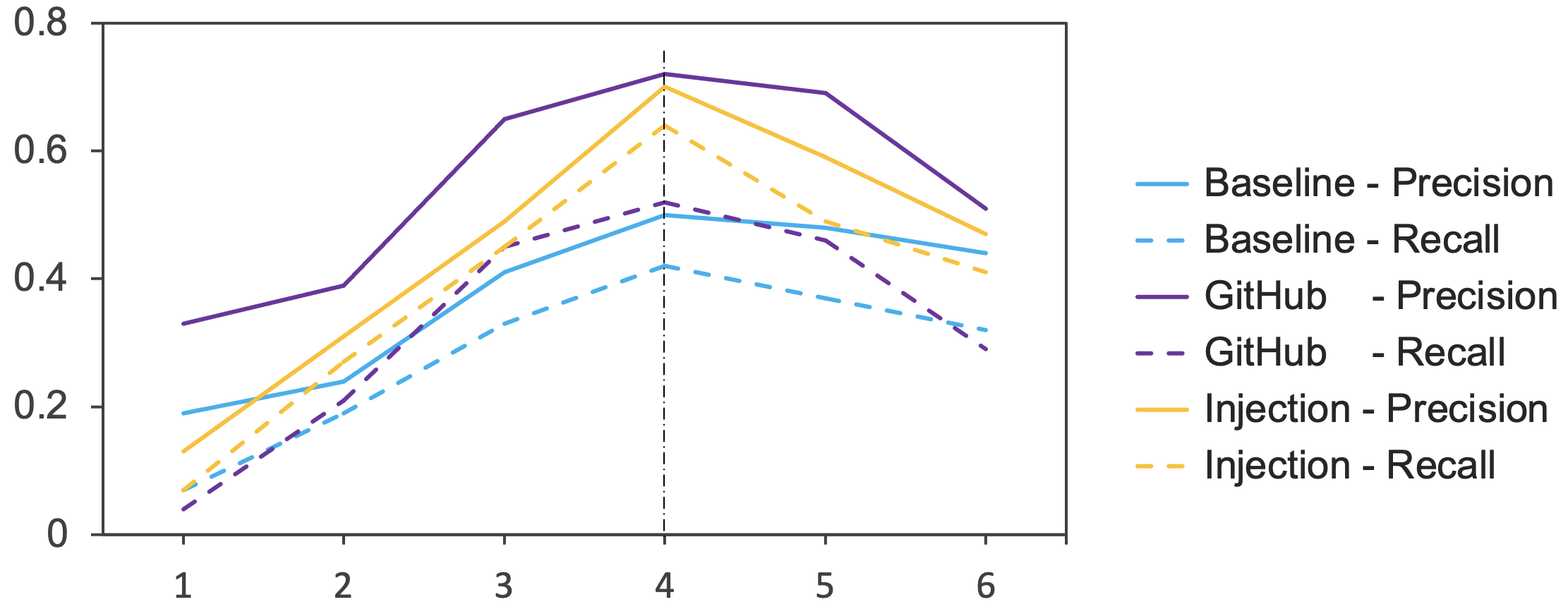}
\caption{Different number of examples (RQ2)}
\label{fig:RQ2-2}
\vspace{-0.1in}
\end{figure}

\subsubsection{\textbf{Influence of Different Number of Examples}}
\label{subsub_results_RQ2_Main_Example_data}

Fig. \ref{fig:RQ2-2} demonstrates the performance under the different number of examples provided in the prompt. 
We can see that the non-crash bug detection performance increases with more examples, reaching the highest bug detection rate with 4 examples.  
And after that, the performance would gradually decrease even increasing the examples. 
This indicates that too few or too many examples would damage the performance, because of the tiny information or the noise in the provided examples. 

\begin{table}[htb]
\vspace{-0.05in}
\renewcommand\arraystretch{0.8} 
\caption{Information of the fixed bugs. (RQ3)}
\vspace{-0.05in}
\label{tab:RQ3-Usefulness}
\centering
\scriptsize
\begin{tabular}{p{0.55cm}<{\centering} | p{1.9cm}<{\centering} | p{1.2cm}<{\centering} | p{1.3cm}<{\centering} | p{1.3cm}<{\centering}}
\toprule
\textbf{Id} & \textbf{APP Name} & \textbf{Download} & \textbf{Category} & \textbf{Version}\\
\midrule
1 & Waze & 500M+ & map & 4.103 \\ 
\rowcolor{gray!15}
2 & MEGA & 100M+ & productivity & 13.5\\ 
3 & Swiggy & 100M+ & food & 4.61\\ 
\rowcolor{gray!15}
4 & AlfredCamera & 50M+ & house & 2024.2\\ 
5 & LG ThinQ & 50M+ & lifestyle & 5.0.2\\ 
\rowcolor{gray!15}
6 & SmartNews & 50M+ & news & 24.7.55\\ 
7 & Da Fit & 50M+ & sport & 2.7.7\\ 
\rowcolor{gray!15}
8 & Blinkit & 50M+ & food & 16.11.0\\ 
9 & Bose & 5M+ & music & 10.2.4\\ 
\rowcolor{gray!15}
10 & TomTom & 5M+ & map & 9.51\\ 
11 & ExpressPlus & 5M+ & medical & 4.12.0\\ 
\rowcolor{gray!15}
12 & Property & 5M+ & house & 5.267.0\\ 
13 & Plant Parent & 5M+ & lifestyle & 1.76\\ 
\rowcolor{gray!15}
14 & NOS & 5M+ & news & 202405\\ 
15 & CommBank & 10M+ & finance & 5.11.0\\ 
\rowcolor{gray!15}
16 & AIMP & 10M+ & music & 4.1\\ 
17 & Wise & 10M+ & finance & 8.71\\ 
\rowcolor{gray!15}
18 & Smart Life & 10M+ & lifestyle & 5.15.1\\ 
19 & Zepp & 10M+ & sport & 8.11\\ 
\rowcolor{gray!15}
20 & Hungerstation & 10M+ & food & 8.0.1\\ 
21 & Lumosity & 10M+ & education & 2024.02\\
\rowcolor{gray!15}
22 & Canvas Student & 10M+ & education & 7.5.0\\ 
23 & Khan Academy & 10M+ & education & 8.1.1\\ 
\rowcolor{gray!15}
24 & Healthengine & 1M+ & health & 10.1.11\\ 
25 & Virgin Australia & 1M+ & travel & 2.33.0\\ 
\rowcolor{gray!15}
26 & Opal Travel & 1M+ & travel & 9.8.2\\ 
27 & NAB Banking & 1M+ & finance & 9.153.0\\ 
\rowcolor{gray!15}
28 & Flybuys & 1M+ & shopping & 24.7.1\\ 
29 & EverydayRewards & 1M+ & shopping & 24.11.0\\ 
\rowcolor{gray!15}
30 & Woolworths & 1M+ & shopping & 24.14.0\\ 
31 & Home Assistant & 1M+ & house & 2024.5\\

\bottomrule

\end{tabular}
\vspace{-0.15in}
\end{table}

\subsection{\textbf{Usefulness Evaluation (RQ3)}}
\label{sec_results_RQ4}
For the 187 apps, {\tool} detects 102 non-crash bugs in 74 apps, of which 43 bugs in 42 apps are never discovered before. 
Furthermore, only 9 of these new bugs are detected by Genie~\cite {su2021fully} and Odin~\cite{wang2022detecting}.
We submitted these 43 bugs to developers, and all of them have been fixed/confirmed so far (31 fixed and 12 confirmed, none of them rejected). 
This further indicates the effectiveness of our proposed {\tool} in bug detection.
Due to space limit, Table \ref{tab:RQ3-Usefulness} presents a part of fixed/confirmed bugs, and the full lists can be found on our website\textsuperscript{\ref{github}}.
We also collect the feedback from developers regarding the bugs we submitted.``This bug is crucial because it affects a core functionality that users frequently interact with.'', ``We’re aware that this bug is hindering users' ability. It’s a high priority for us to address this issue.''. These feedbacks also demonstrate the importance of {\tool} in detecting non-crash bugs.

\section{Discussion}
\label{sec_discussion}
This Section further discusses the generality of {\tool} across platforms.
Almost the existing studies of non-crash bug detection~\cite{guo2022detecting, sun2023characterizing,su2021fully,wang2022detecting,farooq2020livedroid} are designed for a specific platform (Android), which limits its applicability in real-world practice.
In comparison, the primary idea of {\tool} is to explore the app and detect the non-crash functional bugs with a visual understanding of MLLM.
Since the screenshots from different platforms (e.g., Android, iOS, Harmony OS) exert almost no difference, {\tool} can be generalized for App exploration and bug detection in other platforms. 
For platforms like iOS and Harmony OS, where text extraction might not be as reliable or feasible, we adapt {\tool} by deleting the textual GUI information from the prompts, and using the remaining modules of {\tool} to guide the app exploration and bug detection process.

\begin{table}[htb]
\renewcommand\arraystretch{0.9} 
\caption{Cross platform performance of {\tool}}
\vspace{-0.05in}
\label{tab:Cross platform}
\centering
\footnotesize
\begin{tabular}{p{1.0cm}<{\centering} | p{1.8cm}<{\centering} | p{1.5cm}<{\centering} | p{1.2cm}<{\centering} | p{1.2cm}<{\centering} }
\toprule
\textbf{Platform} & \textbf{Code coverage} & \textbf{Report bug} & \textbf{Real bug} & \textbf{Precision}\\ 
\midrule
IOS & 0.47 & 15 & 9   & 0.60 \\ 
Harmony & 0.57 & 19 & 13 & 0.69 \\ 
\bottomrule
\end{tabular}
\end{table}

We conduct a small-scale experiment for the popular platform iOS and Harmony OS, and collect 10 open-source apps for testing on each platform. 
As shown in Table \ref{tab:Cross platform}, results show that the average code coverage is 0.47 in IOS and 0.57 in Harmony OS. The bug detection precision in IOS is 0.60 and the precision in Harmony OS is 0.60. 
This is comparable to the performance in Android, and further demonstrates the generality of {\tool}. We will conduct more thorough experiments in the future.

\section{Related Work}
\label{sec_related}
\subsection{\textbf{Automated GUI Testing}}
To ensure the quality of mobile apps, many researchers study the automatic generation of large-scale test scripts to test apps~\cite{xie2007designing}.
Since Android apps are event-based~\cite{anand2012automated,wu2019analyses,jabbarvand2019search,matinnejad2017automated}, the most common automated testing methods are model-based automated GUI testing methods~\cite{mirzaei2016reducing,yang2018static,yang2013grey,zeng2016automated,mao2016sapienz,su2017guided,dong2020time,gu2019practical,wang2020combodroid}, design corresponding models through the analysis of the apps.
Due to the lack of consideration for the semantic information of the app's GUI pages, the coverage of model-based methods is still low.
Researchers further proposed human-like testing strategies and designed learning-based automated GUI testing methods~\cite{li2019humanoid,pan2020reinforcement}. 
GPTDroid~\cite{liu2023chatting} used GPT-3.5 to generate the testing script, aiming at proposing a more effective approach to generate human-like actions for testing the app. 
AppAgent~\cite{yang2023appagent} is a novel LLM-based multimodal agent framework, that enables the agent to operate apps through a simplified action space.
Qtypist~\cite{liu2023fill} and AudoDroid~\cite{wen2024autodroid} combine LLM and traditional GUI testing tools during the exploration process.
However, these methods require first converting the GUI visual information into text, which can cause significant visual information loss during the process. Moreover, a large number of app widgets and pages are generated through dynamic rendering, and the attribute information of image widgets will be lost when obtaining view hierarchy files. As a result, the coverage of automated testing is low, and many widgets and operations are missed.
This study employs the MLLM for the GUI testing, which is more effective through aligning the GUI screenshot with text information for the vision prompt.

\subsection{\textbf{Non-crash Functional Bug Detection}}
According to the sources of test oracle, current research can be divided into four types. The first type utilizes oracle provided by testers~\cite{adamsen2015systematic} to find non-crash bugs in different strategies (such as ChimpCheck~\cite{lam2017chimpcheck}, AppFlow~\cite{hu2018appflow}, ACAT~\cite{rosenfeld2018automation}, AppTestMigrator~\cite{behrang2019test}, CraftDroid~\cite{lin2019test}). 
Unlike them, this paper doesn't require manual direct/indirect oracle.

The second type uses differential testing to overcome the problem of missing oracles. 
Genie~\cite{su2021fully} and Odin~\cite{wang2022detecting} used the metamorphic relation of independent view property to find bugs.
SPAG-C~\cite{lin2014accuracy}, DiffDroid~\cite{fazzini2017automated} and dVermin~\cite{su2022metamorphosis} compare screenshots between two different mobile devices or platform versions to identify bugs. 

The third type is the automatic generation of test oracle for specific categories of user interactions. 
ITDroid~\cite{escobar2020empirical} compared the GUI pages of an app under two different language settings to find i18n issues.
SetDroid~\cite{sun2023characterizing} generated the oracle for system settings bugs, but can't detect other types of bugs.

The fourth type is to use labeled large-scale bug data to train computer vision models for detecting bugs related to GUI display. The OwlEyes~\cite{liu2020owl}, Gilb~\cite{chen2021glib} and NightHawk~\cite{liu2022Nighthawk} are based on large-scale labeled UI bug data, to train computer vision models that can be used to detect the display bugs.

However, these bug detection methods have problems such as a single category of bug detection, requiring a large amount of manually labeled test oracle data, and being unable to meet complex business scenarios and app functionality.


\section{Conclusion}
\label{sec_conclusion}
With the advancement of LLM, there have been tremendous applications of LLMs in software engineering-related tasks such as test case generation with great performance improvement.
Subsequently, MLLMs emerge and are applied in image-text reasoning, video generation, etc., further expanding our understanding of the capabilities of LLMs. 
this paper proposes {\tool}, a novel vision-driven, multi-agent collaborative automated GUI testing approach for detecting non-crash functional bugs. It comprises three agents: Explorer, Monitor, and Detector. 
This work contributes as a stepping stone for future exploration in software engineering, offering insights and guidance on the potential applications of MLLM. 

In the future, we will further improve this approach for better performance.
Meanwhile, beyond Android apps, we will customize and apply our approach to other platforms such as web, desktop and automotive OS for software testing.






\bibliographystyle{IEEEtran}


\bibliography{reference}

\end{document}